# Modified Brownian Motion Approach to Modelling Returns Distribution

*Gurjeet Dhesi (dhesig@lsbu.ac.uk)[1]*

*Muhammad Bilal Shakeel (shakeem2@lsbu.ac.uk)*

*Ling Xiao (xiaol4@lsbu.ac.uk)*

## Abstract

An innovative extension of Geometric Brownian Motion model is developed by incorporating a weighting factor and a stochastic function modelled as a mixture of power and trigonometric functions. Simulations based on this Modified Brownian Motion Model with optimal weighting factors selected by goodness of fit tests, substantially outperform the basic Geometric Brownian Motion model in terms of fitting the returns distribution of historic data price indices. Furthermore we attempt to provide an interpretation of the additional stochastic term in relation to irrational behaviour in financial markets and outline the importance of this novel model.

---

[1] School of Business, London South Bank University, SE1 0AA.

**Introduction:**

In traditional financial theory with respect to the efficient market hypothesis the main assumption is about the rational expectations of investors with respect to future prices of assets/shares which are assumed to reflect all available information (Fama, 1965, 1970). Accordingly financial theories and models assume that continuously compounded financial returns are normally distributed. However empirical evidence suggests that many financial returns series are leptokurtic. It is well known that many different approaches and methodologies have been applied to investigate the market behaviour, pricing process and to model returns distributions. These include the early approach of Mandelbrot (1963a, 1963b), the employment of stable distributions and jump diffusion models. Extensive literature exists on this; the reader is referred to Mills & Markellos (2008: chapter 7), Rachev et al (2005: chapters 3 and 7) and Birge & Linetsky(2008: chapters 2 and 3) and references therein contained.

The basics of quantitative finance still rely heavily in line with rational behaviour of investors and weak form EMH. That is continuous financial returns can be expressed as

$$r_t = \ln(\frac{P_t}{P_{t-1}}) = \mu + \varepsilon_t \qquad \varepsilon_t \sim NID(0, \sigma^2) \qquad (1)$$

where μ is the average returns and $\varepsilon_t$ is assumed to be normally independently distributed with zero mean and constant variance. The above equation can be written as

$$P_{t+\delta t} = P_t \exp(\mu \delta t + \sigma Z_t \sqrt{\delta t}) \qquad (2)$$

where $Z_t$ is a random number drawn from standardised normal distribution and $\delta t$ is a small time step. This equation is deployed to run simulations and construct the modelled returns distribution based on the Geometric Brownian Motion (GBM).Furthermore the continuous time version of the above equation is

$$\frac{dP}{P} = \exp(\mu dt + \sigma Z \sqrt{dt}) - 1 \qquad (3)$$

Applying Ito's Lemma the equivalent stochastic differential equation (SDE) form of (3) is expressed as

$$\frac{dP}{P} = \alpha dt + \sigma Z \sqrt{dt} \quad \text{where } \alpha = \mu + \frac{1}{2}\sigma^2 \qquad (4)$$

The above model (equations (1), (2), (3) & (4)) provides the foundations of classical quantitative finance and further financial modelling rely on this representation. As mentioned above the problem is that the distribution of returns generated from this GBM model does not match the distribution of historic returns data which often show leptokurtosis. This paper aims to modify the GBM model for financial returns modelling. The motivation came from an experimental paper (Dhesi et.al. (2011), *this paper is attached as appendix A*) about semi closed stock market. The model used for the experiment was the GBM model with extra factors of demand and supply embedded into it.

This study enhances the matching power of the GBM to historical returns distribution by adding a function of Z multiplied by the mean and a parameter $K$. Of course when $K = 0$ we recover the GBM. In line with $Z_t$ being an innovation at time t it can also loosely be considered as news generated at time t. The modified model in its general form is

$$P_{t+\delta t} = P_t \exp(\mu \delta t + \sigma Z_t \sqrt{\delta t} + \mu K f(Z_t) \delta t) \qquad (5)$$

The model specified in equation (5) will now be referred to as the Modified Brownian Motion Model (MBMM). It may also be referred to as the Stochastic Mean Model.

This modified specification is **important** as this endogenously generates (not a theoretical exogenously imposed distribution) a distribution, when choosing the appropriate realisation of $f(Z)$, that is leptokurtic and hence is appropriate for return distribution. Also it can be noticed that only the normally distributed Z innovation appears in this model compared to the extensive literature on modelling returns distributions using jump diffusion processes which contain normal and Poisson jumps. This shows that the MBMM is more parsimonious and accessible. This modelling process can also be seen as a move from general modelling (GBM) to specific modelling (MBMM) of specific return data sets in specific time periods.

Furthermore this modelling process can supply tentative links to irrational behaviour in finance and also show how contribution to knowledge can be gained in the context of providing forecast models of exceedence correlation between returns and copula theory and application enhancement and subsequent implications to portfolio optimisation. Justification of this paragraph is supplied in the Further Discussion section.

**Analysis and Results:**

The realisation of $f(Z)$ is dependent on extensive empirical analysis of historical data. The function we propose is:

$$f(Z) = (2\exp(-c\frac{Z^2}{2}) - 1)\arctan(Z) \qquad (6)$$

The analysis was performed on various market indices. The software package employed for extensive simulations is MATLAB. Fit for purpose optimal value of parameters $K \ and \ c$ for specific data sets are selected by chi-squared goodness of fit statistics.

For illustration purposes: Figure (1) below provides a comparison of the returns distribution of Historic data (histogram), GBM (green) and MBBM (red). Daily data set S&P500: 1st Jan.2010- 31st Dec.2011

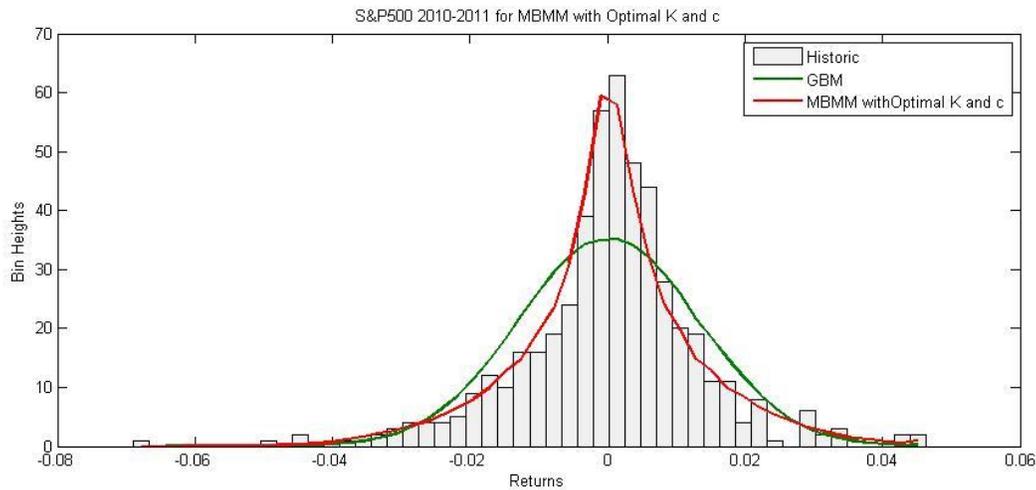

Fig(1): S&P500 with GBM and MBMM (K=-28,c=0.6)

It can be seen in the above figure that the red curve MBMM (with $K_{optimal} = -28, c_{optimal} = 0.6$) is very close to the historic histogram as compared to the GBM represented by the green curve. This was further verified by running a chi-square goodness of fit test on historic data (frequency observed) with simulations from the GBM and the MBMM (corresponding expected frequencies).

The historic data is represented into a histogram and the frequency of each bin as observed frequency is noted. Then we model the corresponding expected frequencies from the GBM and MBMM on same bins by running extensive simulations and taking average of these bin heights of these simulations as expected frequency. Limits of $\mu \pm 3\sigma$ are considered and the bins are customised such that each bin has a frequency of at least 1% of total values in data set. The adjusted frequencies are labelled as customised observed frequency ($f_{oc}$) for historic data and customised expected frequency ($f_{ec}$) for corresponding GBM and MBMM. The goodness of fit statistics for the GBM and the MBMM are calculated by $\chi^2 = \sum \frac{(f_{oc}-f_{ec})^2}{f_{ec}}$ with $(n - p - 1)$ degrees of freedom. Where $n$ is the number of customised bins in data set and $p$ is the number of parameters being estimated for the data set.

Applying the above mentioned technique, the customised distribution (21 bins in customised distribution) has chi-square total of 77.04 (p value=2.80E-09) for GBM and 14.93 (p value=0.53) for MBMM. Therefore we can infer that the MBMM provides a much better fit to the historical data. Furthermore Kurtosis for the customised historical data is 4.06 and the modelled Kurtosis from MBMM is 3.83

The summary results for MBMM for the subsequent two year time period i.e. daily S&P500 returns from 1st Jan.2012 to 31st Dec.2013are also presented here. The chi-square total for this data (26 bins in customised distribution) is 32.57 (p value=0.09) for

GBM and 19.97 (p value=0.52) for MBMM ($K_{optimal} = -3.5$ & $c_{optimal} = 0.6$). Kurtosis for the customised historical data is 3.50 and the modelled Kurtosis from MBMM is 3.57.

Naïve forecasting process may be based on using $K_{optimal}$ and $c_{optimal}$ of the current time series returns distribution for the subsequent time period returns distribution, this will be an improvement on the GBM model, further research is in progress to improve the forecasting mechanism.

We also applied the modified model on many other indices and also for different time periods. One notable aspect of this exercise is that optimal $K$ $and$ $c$ values are different for each data set because assets for different time periods and from different markets possess different characteristics. Price movement of assets are influenced by different sets of financial, political and economic factors Cont (2000). Thus MBMM parameters change for different data sets. This is in line that if we consider GBM as general modelling process then the MBMM is specific modelling process. The goodness of fit chi squared value (for optimal $K$ $and$ $c$) for most of the cases gives a p value>0.05.

For diagnostic purposes: the MBMM was applied on daily data as shown in the illustrative study; however the results lead to the question that what will be the outcome if extra function of Z is multiplied by σ instead of μ. The MBMM model was applied on 2 day data (μ and σ still kept at the daily level and as now $dt = 2 \Rightarrow \sqrt{dt} = 1.4$. The resulting simulation outcome was consistent with those from daily data when $f(z)$ was multiplied with mean, however complete lack of fit and inconsistency when $f(z)$ was multiplied with sigma, confirming the MBMM as a stochastic mean model.

Moreover, the MBMM was transformed into the equivalent SDE version by applying Ito's Lemma.

$$\frac{dP}{P} = \alpha dt + \sigma Z\sqrt{dt} + \alpha K f(Z)dt \quad \text{where} \quad \alpha = \mu + \frac{1}{2}\sigma^2 \qquad (7)$$

This may not be in full rigour however this is applicable when μdt is small, as is verified when extensive simulations were run using the discrete form version of equation (7) and the results were found to be consistent to those obtained from the exponential form (equation (5)) of MBMM.

The MBMM was also applied on larger daily data set to analyse the performance. The summary results of the 10 year daily data from S&P500 index from 1st Jan.1994 to 31st December 2003 are presented here. This data (75 bins in customised distribution) has chi-square total of 251.37 (p value= 1.04E-21) for GBM and 90.68 (p value=0.05) for MBMM ($K_{optimal} = -10$ & $c_{optimal} = 0.4$). Kurtosis for the customised historical data is 3.78 and the modelled Kurtosis from MBMM is 3.64.

Furthermore the MBMM was applied on subsequent ten year (1st Jan.2004 to 31st December 2013) turbulent time period to observe the output for optimal values of K and c. This data (50 bins in customised distribution) has chi-squared value of 614.26 (p-value=7.6E-100) for GBM and 88.4(p-value= 0.00012) for MBMM ($K_{optimal} = -30.8$ & $c_{optimal} = 0.2$). Kurtosis for the customised historical data is 4.95 and the modelled Kurtosis from MBMM is 4.32. A significant reduction in chi-squared value (as can be seen by huge improvement in p-value) indicates that MBMM provides a closer fit to historic data as compared to GBM. Further research is being carried out on how to model highly turbulent time series returns distributions.

Modelling behaviour of MBMM on big (i.e. fifty year) monthly data sets was tested and it is encouraging to see that here again MBMM provides a superior fit. The optimal value of $K$ is again negative with the absolute size being smaller as may be expected.

**Further Discussion:**

The continuous and differentiable function $f(Z)$ for different negative values of $K$ is plotted below. For illustration purposes arbitrarily we have plotted the function for $c = 1$.

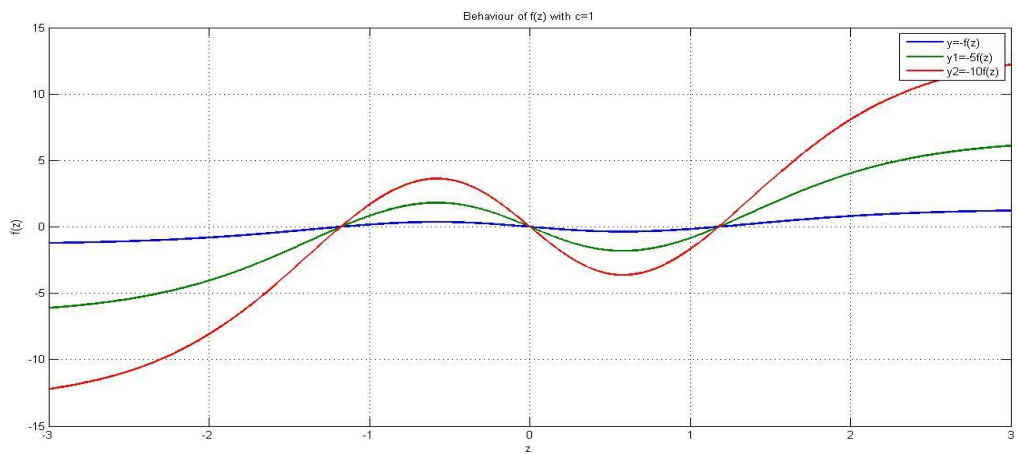

Fig (2): Kf (Z) for different negative values of K. c=1.

The roots of this function depend on the value of the positive parameter c. The bigger the value of c the closer the roots are to the origin. Hence c is the controlling parameter where the modelling the flattening of the tails starts in terms of the number of standard deviation away from the mean. For example if c=1 then the flattening of the tails starts at 1.18 standard deviations away from the mean value. Below a snapshot of the positive tail (empirical, GBM, MBMM) is provided for a specific data set. It can be seen that the

MBMM provides a close fit to the **fat** empirical tail.

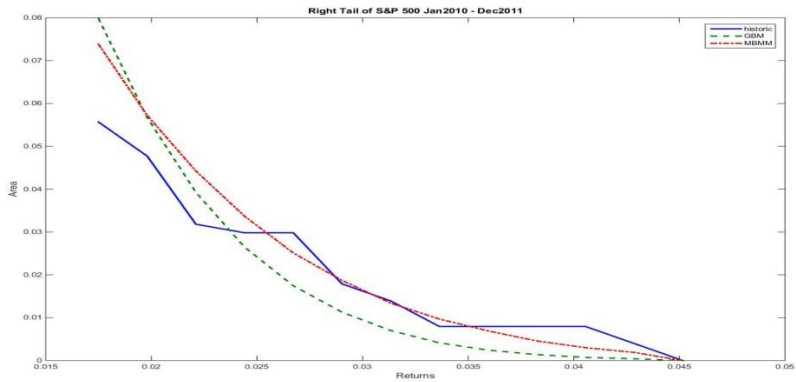

Fig(3).Right Tail of S&P500 for Historic, GBM & MBMM

Now, consider *positive* values of Z at time t (explanation for negative values of Z will follow in a similar but opposite manner). When $Z_t$ is small (small in this case implies: $0 < Z_t < Z_{positiveroot}$ ), this is up to the positive root of $f(Z)$, the value of $Kf(Z_t) < 0$. This in turn has the effect of $P_{t+\delta t}$ from the MBMM to be smaller than $P_{t+\delta t}$ for the GBM, which in turn peaks the distribution of $r_t$ near the mean value. Now if we allow $Z_t$ that is the innovation to be perceived as news then the following connection can be made: when the news of the upturn in the market is minimal, not substantial, then in aggregate the movers (investors) of the market perceive that there is sluggishness in the market ( not much push in the market towards the positive direction) become impatient (and hence irrational: call this negative irrationality) and decide to sell the assets that belong to the market, so as to invest their money in alternative products, hence supply outstrip demand and hence the price increase as modelled in the MBMM model diminishes compared to the rational GBM model. Now when $Z_t$ becomes positively large (large in this case implies: $Z_t > Z_{positiveroot}$ ), this is beyond the positive root of $f(Z)$, the value of $Kf(Z_t) > 0$. This in turn has the effect of $P_{t+\delta t}$ from the MBMM to be greater than $P_{t+\delta t}$ for the GBM, which in turn flattens the distribution of $r_t$ in the tail areas. In this case the market movers perceive that the market is positively bullish and in aggregate herd in an irrational behaviour (positive trading irrationality) to buy into the market, demand outstrips supply and hence the price increase as modelled in the MBMM model inflates compared to the rational GBM model. As may be noted from our illustrative study results from table (1), c_optimal=0.6 implies $Z_{root} \cong 1.5$, $Z_{root}$ may be considered as a transition point.

Furthermore the 180 degree rotational symmetry around the origin (hence the roots will no longer be equidistance from the origin) can be broken to model skewness. This is achieved by introducing a parameter m in the exponential part of the $f(Z)$ (and not in the arctan(Z)) i.e $Z^2$ changes to $(Z-m)^2$ and negative or positive skewness can be modelled with appropriate positive or negative values of m respectively.

In essence the extra function $f(Z)$ alongside the parameters $K\ and\ c$ model the irrational behaviour of financial market price indices which is not captured by the weak form EMH. In line with general to specific modelling we propose that equations (5), (6) and (7), a parsimonious and accessible modification, be considered as the basis for modelling specific returns distributions in specific time periods with the appropriate optimal parameters.

Although we believe that the MBMM may not change the Black Scholes Option Pricing formula, the MBMM will significantly contribute to other areas of Finance, a couple of areas where this applies are outlined below:

Application in Exceedance Correlation: Longin and Solnik (2001) advocate the use of the 'exceedance correlation' to examine the asymmetry in correlation in extreme situations between different stock markets. 'Exceedance' refers to the extreme values of the asset distribution and it is defined as values exceeding certain confidence boundries, like 95 percent of the distribution. However, the exceedance correlation so far is only useful to testing the asymmetric correlation for a given empirical period. With the MBMM proposed in this paper, we may short term forecast the future exceedance correlation i.e. we can run simulations for next t time period price using Equation (5) with the optimal K, c value. Then group the simulated values into different bins and compute the group correlations according to:

$$\tilde{\rho}(q_\alpha) \equiv \begin{cases} \text{correlation } [X, Y | X \leq Q_x(q_\alpha) \cap Y \leq Q_y(q_\alpha)], \text{for} q_\alpha \leq 0.5 \\ \text{correlation } [X, Y | X > Q_x(q_\alpha) \cap Y > Q_y(q_\alpha)], \text{for} q_\alpha \geq 0.5 \end{cases}$$

Application in Portfolio Optimisation: Mendes and Marques (2012) argue that a better estimation of the portfolio efficient frontier depends on a reliable characterisation of the data underlying multivariate distribution. This requires a proper modelling of the individual assets as a critical step in portfolio management. Authors such as Patton (2004) demonstrate that higher moments improve the portfolio allocation if returns are distributed in a very different way from Normal. Higher moments measurements describe irrational behaviour in financial markets. Patton (2004) proves that higher moments such as skewness and kurtosis significantly affect portfolio allocation and, furthermore, the asymmetry in marginal distribution also affects portfolio management. The Modified Brownian Motion proposed in this paper better reflect the real distribution of assets returns and also successfully provide an interpretation in relation to irrational behaviour in financial markets. Therefore, the MBMM could promote both the theory and practice of asset pricing and portfolio investment.

The next big question is just as eq.(1) (ARIMA (0,0,0)) is the financial econometric discrete version of GBM model as described in eq.(3) and (4): what is the financial econometric discretised version of the MBMM presented in eq. (5) and (6)? Answering this would be very fruitful as it would help us understand this model further and be very helpful for financial applications; a couple of these applications are outlined above (exceedance correlation and portfolio optimisation).

# Appendix A

**Semi-Closed Simulated Stock Market: An investigation of its components**

*Gurjeet Dhesi*

*Email: dhesig@lsbu.ac.uk*

*Mohammad Abdul Washad Emambocus*

*Email: emambocus@lsbu.ac.uk*

*Muhammad Bilal Shakeel*

*Email: shakeem2@lsbu.ac.uk*

*Abstract*

*This paper aims at designing the different important components of a semi-closed simulated stock market (pricing mechanism, stock allocation and news generation). The purpose is to understand the interactions of the different aspects within a 'semi-closed' system. The complexity and nature of the system led to the process of modifying the pricing mechanism which is viewed from a different angle to the classical Brownian Motion and the Random Walk model. However, it incorporates the essence of these two fundamental theories and then investigates the matrix of investors' behaviours in relation to news feedbacks. This paper also explores the realm of randomly generated news to the responses of participants to determine rational and irrational behaviours. This is carried out through uncompressing the time within the experiment and looking at concordant and disconcordant behaviour. The focus is on how the modified pricing equation adapts to the conditions and uniqueness surrounding a semi-closed stock market. Thus, this paper looks at how a simple market system where the main determinants of share prices are news, demand and supply along with some filtering of the external forces can affect the behaviours of investors in terms of*

*their portfolio composition. The return distributions can then be stipulated as arising from rational or irrational trajectories and subsequently be simulated and matched via the proposed modified Brownian Motion model to empirical return distributions in specific time periods and markets.*

*JEL classification*: **D03; E37; G12; G14; G17**

*Keywords: Semi-closed stock market; pricing mechanism; modified Brownian motion; news feedbacks; irrational trajectory*

1. Introduction

The perception of financial markets tends to differ considerably between the two worlds of academic theorists and market practitioners. In traditional financial theory such as the efficient market hypothesis (Fama, 1970), the main assumption is about the rational expectations of investors in respect to future prices of assets/shares which are assumed to reflect all available information. This implies less opportunity to incur speculative profit generated from the application of technical trading. Thus, changes in the market such as price overreaction (bubble and bursts) are considered as aspects of rational changes in the valuation of shares rather than through sudden shifts or drifts in investor's sentiment or mood. Traditional theorists view the market as rational, mechanistic and therefore efficient where trading volume and price volatility are not serially correlated. On the other hand, market practitioners have a different take on the function of the market and on the determinants of prices. They view the market as the main platform for offering speculative opportunities. Technical trading is viewed by market agents as profitable where there is the existence of a market psychology which explains financial anomalies such as herding behaviours of investors. The market is even viewed as a single entity which possesses its own personality, mood and varying sentiments. Agents have described the market as being sluggish or jittery showing that it is not consistently rational in its core as theoretised by traditional economists. The practitioners see the market as being highly organic, psychological and imperfectly efficient in nature where market agents are viewed as human not econs (Thaler and Sunstein, 2008), where their moods influence or determine the health of the market. Consequently, there is a disagreement between theorists and traders which is why the different aspects of asset pricing and the overall market investigation cannot be carried out in a vacuum, both ends should be scrutinised to determine the way forward for a modern finance.

An agent-based financial market such as a stock market is an ideal arena for investigating and understanding behaviours of investors/individuals which differs from the notions of traditional point of view of equilibrium. Various domains such as psychology, physics, economics and computer science are having a growing interest in exploring financial markets and agents' behaviours. Most of the studies (Singal, 2004, Frey and Eichenberger, 2002 and Shleifer, 2000) conducted around examining the behaviours of investors were done empirically through the analysis of the financial market and therefore explaining the different anomalies. These "hiccups" have been the main determinant or trigger to serious economic and financial upheavals during the last century. The focus of researchers have been more driven away from a macro level and stressing the importance at investigating the micro-structural aspects of the financial market in order to clearly paint a true picture of the whole financial world. Along these lines of thoughts, studies (Farmer and Joshi, 2002 and Takahashi and Terano, 2003) have been carried out at dissecting the various components of stock markets; these involve the behaviours of agents, the pricing mechanism and the allocation problems that determine the main structure or spine of the financial sphere.

Market microstructure focuses on the detailed mechanics and functioning of the financial markets. It aims at analysing the interconnection of the impacts of the market on prices of assets, volume and the trading behaviour (De Fong and Rindi, 2009). Theories, such as the Walrasian school of thought, stipulate that in perfect markets the Walrasian equilibrium price encapsulates the demand curves of all potential agents in the market. This determination of the equilibrium in the market looking at aggregated demand and supply is aimed at determining the market price. This system clearly looks at how the demand of investors is translated into prices and volumes which enables the financial market to operate.

The simple model of security prices provides a basic and straightforward way of tackling the issue of asset pricing. This equation is given in an arithmetic version which is used as a reference for designing the pricing model for the experiment in this paper. This model assumes weak form market efficiency where prices reflect all public information. This means that if investors are assumed to have symmetric information and where frictions are negligible then prices reflect the expected values. The discrete returns are provided by $\frac{P_t - P_{t-1}}{P_{t-1}}$ and the mean value is specified as $\mu_a$. The continuously compounded returns are the logarithmic difference of the price relative and the mean return is specified as $\mu_g$. When modelling the returns as a white noise process, the random error denoted as $\varepsilon_t$ is added onto the mean return. $\varepsilon_t$ is assumed to be normally independently distributed or independently identical distributed. This indicates that markets are efficient where prices reflect expected values at all points in time. However, some studies (Oliver, 1989) have demonstrated that the above does not account for bubbles or bursts in the market which means that it does not capture the various behavioural or financial anomalies. This is the main reason why theoreticians have developed sophisticated financial econometric models in order to explain these factors through the investigation of volatility transmission, collective phenomena and herding behaviour. Most of the recent studies in behavioral finance for instance, target at analysing these anomalies present in the distribution of returns (DeLong *et al.*, 1991, McLeavey, and Rhee, 1995, Daniel *et al.*, 1998, Dittrich, Güth and Maciejovsky, 2001) which look at the works carried out in psychology on patterns of human behaviour. Various studies have scrutinised the impact of investors' behaviours on the security prices and the process through which investors update their beliefs in light of new information. However, much of the literature has not completely reviewed how the pricing mechanism absorbs the movements of investor's behaviour. The study of DeLong *et al.* (1991) analyses the impact of noise traders in the financial markets. The findings point out that the irrational behaviours of noise traders are the main instigators of price volatility in the market which acts for the advantage of other groups of investors. Thus, it is acknowledged that irrational behaviours determine the fluctuations in the movements of prices away from the fundamental/intrinsic value (DeBondt and Thaler, 1987).

Following critiques of the simple model's assumption that financial markets operate efficiently where all information is publicly available, new models of asset pricing have been developed in order to reflect a more realistic view of the market. As such, the real financial world operates in a structure with the main determinant of asymmetric information which suggests that there are different types of agents in the market with different access to information. The advances in the literature of market microstructure link the new developments in areas encompassing economics of information, rational expectations and imperfect competition to determine the impact of information and how investors update their beliefs accordingly. Thus, the different categories of investors have different motives within the system. Bagehot (1971) distinguishes between those agents that are motivated by liquidity who do not have any private information over others and informed investors with the advantage of private information. Noise traders are classified as being liquidity driven who adjust their portfolios accordingly to areas where they have the perception that they possess current information. The importance or determinant of information asymmetry allows informed traders to benefit from their private information over uninformed agents. This means that a market maker is on average in a less beneficial position compared to informed traders. However this is compensated by advantages market traders have over noise traders which indicates that the spread pertains to an informational component that shapes the market trajectory (price formation). Kyle (1985) provides a model with a single agent who exercises control over information. The trader fully exploits this advantage of information asymmetry before the information is diluted and becomes publicly available. The market maker examines the net order flow in order to determine the appropriate price which represents the expected value of the asset. According to this model, price is determined after orders are placed. Here only market makers are allowed as opposed to the real financial world where traders can set their demands on price. In this model, Kyle shows that there is a rational expectation equilibrium whereby the market prices eventually reflect all available information. Given that the model assumes a normal distribution and continuous quantities of order taking any value over the real line, this framework has a linear regression. Hence, the price at any point in time is the expected value of the asset having a linear projection. This model shows that the market makers merely perform the duty of an order processor, determining the market clearing prices. It is clear that this model is an appropriate extension or enhancement of the simple model; however there are still some important aspects of the financial market that it cannot fully capture. This also involves analysing for example the 'fat tails' from the returns of securities where volatility is transmitted or generated. The discussion presented in this section leads the paper to focus on modifying the Brownian Motion in order to determine a pricing equation that captures these aspects which are lost within an open market situation, thus showing the importance of running experiments within a semi-closed market.

## 2. Structure and parameters

The process of the research is on a semi-closed laboratory experiment where the stock market environment is being simulated in order to understand the mechanism of pricing formation, news, time factors and behaviours of investors. The different components of this research experiment are elaborated and explained in the following section:

*2.1. Stock Market:*

The experiment is about simulating the situation of a trading environment which involves the buying and selling of particular assets. The premise of this experiment is to design and simulate the situation of a stock market where agents are asked to buy and sell shares depending on the criteria of the system (demand and supply functions and external factors which are filtered into the system). The simulated stock market (SSM) is agent focused or led which means that the experiment is run without any bias or intervention element. The data are generated endogenously through the process of demand and supply (internal forces) and through some external factors. The market consists of *N* heterogeneous agents, who invest inside the semi-closed market. The composition of the SSM is fairly straightforward and simple as the amount of tradable assets have been limited to only four which comprises of four different sectors representing the whole market. The sectors and market is assumed to be fictitious which means that the agents should not allow the situation in the real market affect their decisions. The sectors are Banking, IT/Telecommunication, Retailing and Industrial. These sectors where not broken down into different companies but were just mentioned to be indices with the prices representing the actual soundness of the sectors. The agents were provided with 1000 shares in each of the different sectors and also £20,000 as initial capital for them to invest in the market. The agents need to submit their order form within a trading window which is between 8am till 3pm (British time).

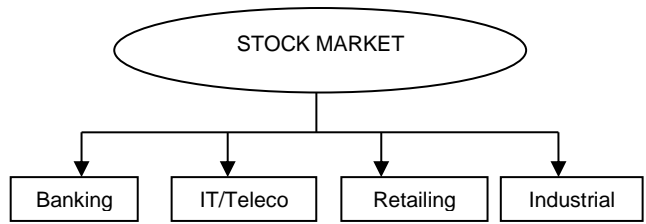

*Semi-closed Experiment:* The agents are introduced to a semi-closed system which means that they need to deal with the internal and external environments. However, it is not an entirely open system which means that the external determinants/forces are being limited under this experiment. The metaphor to explain the nature of this experiment is in terms of an *"inflated balloon"*. The various aspects inside the balloon can affect the shape of the balloon and forces from outside can also determine the composition of the internal factors.

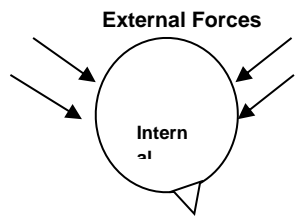

*Trading:* The trading system is through the demand and supply of shares of the different sectors. These are in the form of the shares available to the agents (adjusted daily) and also to their cash flow (remaining from previous trading). Moreover, the agents are not allowed to short sell within the system and allocation of shares is not determined on Bid/Ask limit prices. The prices of the sector shares are computed on the basis of a demand and supply mechanism which is derived through adapting the characteristic of Brownian Motion. Advances in market microstructure along with components of the Brownian Motion were used under such a restrictive system in order to devise an adequate model for determining the current prices of the sectors. This aspect of the pricing mechanism is elaborated in the next section (findings and research).

*2.2. Information Criteria:*

The news provided to the investors contains a mixture of two forms of news. The first form of information news was endogenously generated through analysing the pricing mechanism as determined through the demand and supply mechanism and imbalances at time t. The second form of news is exogenously generated (and provided) by observing the magnitude and sign of the random element drawn from the normal distribution at time t. Hence, the news parameters which are provided to the participants are $netD_{t-1}$ and $Z_t$. These are made available at 5pm on the previous day and before 8am which means before the starts of the new trading session. Note that $Z_t$ is drawn (randomly) between 5pm on the previous trading time and 8am on the trading period. This element acts as news for the participants to consider. It is subsequently entered into the pricing equation to generate $P_{t+1}$.

## 3. Findings and Remarks

*3.1. Experiments:*

Two experiments were conducted where experiment 1 (60 agents participated within this experiment for 15 trading days) acted as a pilot study where the pricing mechanism was analysed and adjusted accordingly. Experiment 2 (20 agents are involved for 35 trading days) has been adjusted from experience gained from the first experience. Data have been collected in order to analyse the performance of the pricing mechanism. The main findings from experiment 1 provided some important insight in the behaviours of agents. It showed that the participants were trading within a falling market where they continue selling the shares which dragged the whole market down. This behaviour is mainly associated to a herding behaviour of investors as they wanted to get out of the market as quickly as possible whereby they thought that selling all their shares will be the quickest way out. However, this frantic behaviour on behalf of the participants led to the crash of the whole market with all the agents incurring losses at the end of the experiment. It is very interesting to analyse how agents within a semi-closed market with no real money or financial risk associated to their trading can still be panicking and creating artificial crashes. This indicates how within the scenario of a falling market agents just wanted to exit the market as quickly as possible without thinking that their actions will further push the market to the abyss. In experiment 2 participants acted in a rational and stable manner and pertained to case (i) in the uncompress time section.

### 3.2. Allocation problem:

As the semi-closed experiment did not have the option of limiting orders with corresponding highest bids and asks rate, the only possible fair allocation mechanism was developed. This is a very simple and straightforward allocation tool whereby the aggregated demand (D) and supply (S) for each sector $i$ at time period $t$ are being matched depending on the minimum total of aggregated demand and supply for each sector $i$ at time period $t$. These variables were allocated with a corresponding ratio of the minimum total and computed accordingly to provide a rounded fair amount to the agents and subsequently spreadsheet updated on a daily basis.

$$\frac{d_{ijt}}{Max[D_{it}, S_{it}]} \times Min[D_{it}, S_{it}] \quad \text{if } D_{it} > S_{it} \quad (1)$$

$$\frac{s_{ijt}}{Max[D_{it}, S_{it}]} \times Min[D_{it}, S_{it}] \quad \text{if } D_{it} < S_{it} \quad (2)$$

$d_{ijt}$ = individual demand of each sector corresponding to agent j at time $t$.

$s_{ijt}$ = individual supply of each sector corresponding to agent j at time $t$.

*Pricing Mechanism:* The main findings from the experiment are the determination of the pricing mechanism which borrows the notion from the Brownian Motion and Random Walk Theory in order to adequately lead to a pricing equation which is applicable to the semi-closed market situation. The different parameters in the system are: the historical growth rate, the general historical volatility of the global market and the transmission coefficient and further to this the aggregate demand (D) in the system, aggregate supply (S) in the system, their adjacent period dynamics and finally a random element $Z$.

The pricing mechanism was in particular highly challenging due to the different specificities of the semi-closed market experiment and the assumptions made for the simulated stock market composition. Most of the theories involved in simulated stock markets base their price determination on variables such as best bid, best offer and time of orders (Cappellini and Ferraris, 2007). However, this experiment mainly deals with the properties of demand and supply from order forms. This allows clear emphasis on analysing the trading behaviours of investors and how they response to the evolution in the market.

*3.3. Stylised Facts:*

The different components and conditions for the pricing mechanism are:

1. The market structure is not a zero-sum market.
2. The price ($P_{t+1}$) for the corresponding time $t+1$ is adjusted specifically to the movements of aggregated demand and supply for the whole market which is a feature of the semi-closed laboratory research.
3. Growth in the system is equal to the growth of the global economy if $netD = 0$.
4. The volatility in the system is equal to the volatility transmission to the system if $\Delta netD = 0$.
5. Transmission coefficient to be determined initially by guestimation.
6. The model is determined with reference to an arithmetic version which is deliberately selected to differentiate between the different components affecting the pricing mechanism[2].

Through the experiment the pricing mechanism was determined and tested for its reliability. After several adjustments carried out within a pilot exercise, the pricing of the shares was determined.

$$P_{it+1} = P_{it}(1 + \mu_m(1 + \frac{netD_{it}}{A_{it}}) + \sigma_{md}(Z_t + \frac{\Delta netD_{it}}{(A_{it} + A_{it-1})/2})) \qquad (3)$$

where:

---

[2] Geometric Brownian motion is given as $P_t = P_{t-1} e^{\mu \delta t + \sigma Z \sqrt{\delta t}}$ which can be transposed arithmetically as $P_t = P_{t-1}(1 + \alpha \delta t + \sigma Z \sqrt{\delta t})$ where $\alpha = \mu + \frac{1}{2}\sigma^2$.

| | |
|---|---|
| $P_{it}$ | = price of a particular sector $i$ at time $t$ |
| $\mu_m$ | = the monthly growth rate of the external market fed inside the system. |
| $netD_{it}$ | = aggregated demand for sector $i$ minus aggregated supply for the same sector at time $t$ |
| $A_{it}$ | = average of aggregated demand and supply for sector $i$ at time $t$ |
| $Z_{it}$ | = a random element (drawn from the normal distribution system) which represent exogenous news provided to the inside sector $i$ of the system at time $t$ |
| $\sigma_{md}$ | = monthly volatility transmission from the external system (the $\sqrt{\delta t}$ and the transmission coefficient is implicitly embedded inside $\sigma_{md}$) |
| $\Delta netD_{it}$ | = the change in $netD_{it}$ between the time $t$ and $t-1$ |
| $(A_{it} + A_{it-1})/2$ | = the average of the addition of the averages of the aggregated demand and supply for sector $i$ at time $t$ and $t-1$ |

This pricing mechanism incorporates the demand and supply factors within the semi-closed system however with no control on behalf of the participants on the way that the allocation is carried out. The pricing mechanism has its main foundations on various important components which are as follows:

### 3.4. Demand and Supply:

There are two demand/supply factors within the pricing equation which reflect the behaviours of participants within the semi-closed system. The first one incorporates the effect of daily demand and supply elements of the sector involved along with a loose normalising factor. The second element relates to the impact of the demand and supply and the changes from previous trading time period with a loose normalising factor that reflect the average aggregated addition of demand and supply for a particular sector. This is an important element within the pricing mechanism as it reflects the behaviours of the participants within the system. Hypothetically randomness associated with prices may affect the volume of demand and supply making them dependent on $Z_{it}$. Therefore, the second factor is added to the $Z_{it}$ which indicates the direct impact of news on the changes of demand and supply.

### 3.5. News and Dynamics:

The generation of the news which is released every time period within the system is established by the impact of the $Z_{it}$ element and the dynamics of the inside trading. A positive value of $Z_{it}$ indicates a positive exogenous news which may rationally or in an exaggerated manner encourage positive $netD_{it}$ especially if the inside investors are prone to the exogenous news and similarly a negative value may drag down the $netD_{it}$ reflecting the situation and reaction of investors due to news that are not favourable for trading. However, this may not be the case as herding behaviour (opposing the exogenous news) may take hold and the news generated from the inside of the system outweighs the $Z_{it}$ element news. Various scenarios regarding the dynamics of the investors and the subsequent price modification of the sectors (and the system) are now possible. Experiment 1 provides some evidences that the majority of the time $netD_{it}$ became negative (positive $Z_{it}$ news did not influence $netD_{it}$ in a positive manner). This led investors to sell in a falling market indicating herding behaviour within the various sectors. However, the main aspect of this herding behaviour is that the trading occurred in a falling market. This clearly indicates that collective phenomenon within the system can have a great influence on the pricing mechanism which may not be in the same line with the parameters or indicators in the system.

### 3.6. Uncompress time into minute factors:

The time uncompression within the pricing mechanism leads to

$$P_{t+1} = (1 + \mu + \sigma Z_{it} + \kappa f(Z_{it}))^3 \qquad (4)$$

where: $f(Z_{it})$ need not be symmetric around the origin when irrational trading is taking place. This leads to the simulation of the pricing mechanism within three distinct trading phenomena – bubble, burst and stable time periods. The time uncompress through simulation results into the opening up of the system (This is equivalent to removing the membrane from our metaphorical "inflated ballon" refered to earlier). Therefore, the three scenarios can now be categorised[4]:

> *In hindsight: scenarios presented here for $f(Z_{it})$ do not provide fit for purpose methods for fitting the empirical returns distribution. The correct $f(Z_{it})$ function is presented in section 4 of the main paper, **"Modified Brownian Motion Approach to Modelling Returns Distribution".***

(i). When $netD_{it}$ over a particular consecutive time (set) period is largely concordant with $Z_{it}$ this reflects the situation of a stable period where the $Z_{it}$ element translates into a different degree (slope) of impact on the dynamics of $netD_{it}$.

---

[3] $\sqrt{\delta t}$ is implicitly incorporated in $\sigma$

[4] These graphs are sketches for illustration purposes and further to this may not necessarily cross the origin.

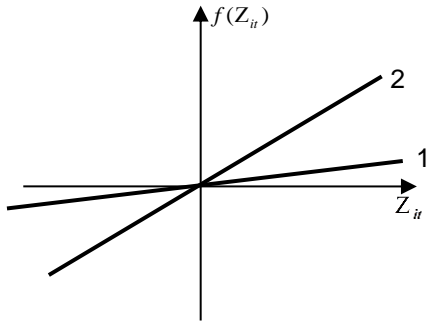

Note that case 2 shows that investors are more sensitive to $Z_{it}$ as opposed to case 1. This is highly dependent on the size of the market, the number of players and liquidity. This can be classified as rational behaviour within the market. This can still be classified as a form of rational behaviour within the market, with the basic random walk model being recovered when the slope become zero.

(ii). When $netD_{it}$ over a particular consecutive time period is disconcordant with $Z_{it}$ within a stable period is quite unlikely to occur within a market because of the element of positive information having a negative impact on $netD_{it}$ over the set time period.

(iii). When $netD_{it}$ is largely concordant with positive news or negative news ($Z_{it}$) but not both this may illustrate periods of bubble or burst, here $netD_{it}$ is likely to follow two different excessive paths within a certain time frame which is not explained through rational behaviour.

(a) 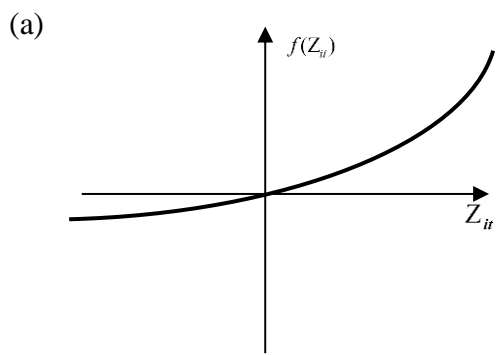

(b) 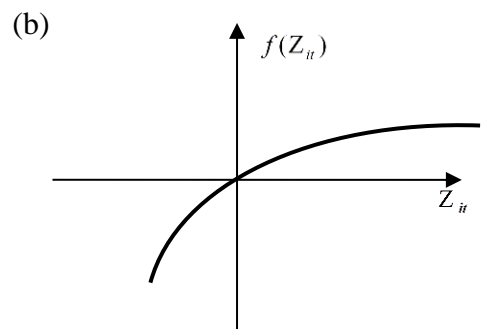

Case (a) reflects a situation of bubble within the market where positive $Z_{it}$ (implying positive news) leads to a positively "significant" $netD_{it}$. However a negative $Z_{it}$ will not have a significant impact on $netD_{it}$. This is of an irrational nature where changes in $netD_{it}$ does not follow the information channels with the system. The market operates as an entity with only one target which is on the overestimated growth of the sectors. Case (b) indicates the reaction of $netD_{it}$ within a period of burst where a negative $Z_{it}$ have a "significant" negative effect on $netD_{it}$ however a positive $Z_{it}$ does not have a significant impact on $netD_{it}$. Case (a) and (b) can be classified as irrational behaviour as concordant behaviour occurs in the first and third quadrant respectively.

## 4. Conclusion and Further Research

Market microstructure is a field which is an aperiodic ebullition with various emphases on determining how irrational trajectory of investors can be linked to the different components of a stock market. Pricing mechanism is one of the main aspects that is examined to see whether it reflects the behaviours of investors. The Brownian Motion which is at the heart of the pricing of shares is being modified to incorporate elements of demand and supply and news feedbacks to reflect the contribution of investors rational and irrational behaviours. The main finding of this paper is the modified Brownian Motion pricing mechanism which has been experimented within a semi-closed stock market in order to capture the behaviours of investors. The pricing mechanism reflects the changes in demand and supply and also the reactions of the participants to the news feedbacks. The second aspect of this paper covers the criteria of news feedbacks and how the system reacts to its movement. Two experiments were carried out with distinct results. Experiment 1 shows that investors disassociated themselves from the news feedbacks and sold their shares with a falling market corresponding to an irrational behaviour. On the other hand, Experiment 2 indicates a more rational behaviour of investors in response to news. Finally, the third component investigated is the time uncompress factor within the semi-closed system. Three main time periods were identified (stable, bubble and burst) in order to reflect the reactions of investors to news. The main hypothesis is that a rational reaction is mainly concordant and disconcordant in nature. On the other hand, periods of bubble and turbulence were classified as either largely concordant with positive news or negative news but not both. Investors disassociate themselves with negative news feedbacks in periods of bubbles and with positive news feedbacks in periods of bursts (evidence of the burst is evident in Experiment 1). Subsequently, news and its effects now becomes part of the model and hence leads to the modified variations of the Brownian Motion as discussed in the previous section.

Further research in terms of the reactions in $netD_{it}$ to changes in $Z_{it}$ provide an adequate platform to run simulations on the model (by investigating the appropriate $f(Z_{it})$ and optimising through the parameter $\kappa$) where the return distributions are matched against real-life empirical return distributions which are generated inside periods of boom, burst and stable times. The simulations will be used to measure the degree of irrational trading in time where the return distributions significantly deviate from the ones generated from the basic random walk model. Note that $Z_{it}$ is the random element in the model, other distributions may correspond; however the $Z_{it}$ factor is hypothesised from the normal random walk model. Furthermore, variations and amendments to the pricing model are being considered. For example, an extra function added to the modified Brownian Motion corresponding to the $\Delta netD_{it}$ arising from the change in news ($\Delta Z_{it}$) over consecutive time period may need to be considered for volatility implications.

The experiments carried out in this paper leading to the modified Brownian Motion will open avenues for further research in the area of collective phenomenon. This is highly related to the news feedback mechanism that investors have been ignoring and following the path of their own irrational trajectory: now to be investigated by identifying the appropriate weight $\kappa$ and the appropriate $f(Z_{it})$ in specific time periods and specific markets.

**Acknowledgement:** We thank Dr Lee Rose, Professor John Warwick and Martin Abram for their valuable comments and suggestions. We express our deep gratitude to Jim Snaith, Head of Business Department at London South Bank University, for all the help provided in the research.